\def\year{2022}\relax
\documentclass[letterpaper]{article} 

\usepackage{aaai23}
\usepackage{times}  
\usepackage{helvet}  
\usepackage{courier}  
\usepackage[hyphens]{url}  
\usepackage{graphicx} 
\usepackage{natbib}  
\usepackage{caption} 
\DeclareCaptionStyle{ruled}{labelfont=normalfont,labelsep=colon,strut=off} 
\usepackage{algorithm}
\usepackage{algorithmic}

\usepackage{newfloat}
\usepackage{listings}

\usepackage{multirow}
\usepackage{setspace}
\usepackage{array}
\usepackage{subfigure}
\usepackage{tikz}
\usepackage{graphicx}
\usepackage{color}
\usepackage{bbding}
\usepackage{amssymb,amsmath,latexsym,amsfonts,amsthm}
\usepackage{graphicx}
\usepackage{booktabs} 
\usepackage{color}

\floatstyle{ruled}
\newfloat{listing}{tb}{lst}{}
\floatname{listing}{Listing}

\title{SDRTV-to-HDRTV Conversion via Spatial-Temporal Feature Fusion}

\author {
    Kepeng Xu\textsuperscript{\rm 1,3},
    Li Xu \textsuperscript{\rm 1},
    Gang He \textsuperscript{\rm *1,2},
    Chang Wu \textsuperscript{\rm 1},
    Zijia Ma \textsuperscript{\rm 1},
    Ming Sun \textsuperscript{\rm 2},
    Yu-Wing Tai \textsuperscript{\rm 2}
}
\affiliations {
    \textsuperscript{\rm 1} Xidian University \\
    \textsuperscript{\rm 2} Kuaishou Technology \\
    \textsuperscript{\rm 3} MEGVII Technology \\
    }

\usepackage{bibentry}

\begin{document}

\maketitle

\begin{abstract}

HDR(High Dynamic Range) video can reproduce realistic scenes more realistically, with a wider gamut and broader brightness range.
HDR video resources are still scarce, and most videos are still stored in SDR (Standard Dynamic Range) format.
Therefore, SDRTV-to-HDRTV Conversion (SDR video to HDR video) can significantly enhance the user's video viewing experience.
Since the correlation between adjacent video frames is very high, the method utilizing the information of multiple frames can improve the quality of the converted HDRTV.
Therefore, we propose a multi-frame fusion neural network \textbf{DSLNet} for SDRTV to HDRTV conversion.
We first propose a dynamic spatial-temporal feature alignment module \textbf{DMFA}, which can align and fuse multi-frame.
Then a novel spatial-temporal feature modulation module \textbf{STFM}, STFM extracts spatial-temporal information of adjacent frames for more accurate feature modulation.
Finally, we design a quality enhancement module \textbf{LKQE} with large kernels, which can enhance the quality of generated HDR videos.
To evaluate the performance of the proposed method, we construct a corresponding multi-frame dataset using HDR video of the HDR10 standard to conduct a comprehensive evaluation of different methods.
The experimental results show that our method obtains state-of-the-art performance.
The dataset and code will be released.

\end{abstract}

\section{Introduction}

\footnote{* corresponding author}

In decades, the technology for producing television content has evolved rapidly.
The resolution of TV content has increased from standard definition to high definition to ultra-high definition. 
The color gamut and dynamic range of TV content have also improved, from BT.709 to BT.2020, from standard dynamic range (SDR) to high dynamic range (HDR).
HDR displays have now appeared in large numbers in everyday life, but corresponding HDR video resources are rare, and most video resources are only available in SDR format.

\begin{figure}[h]
  \centering
  \includegraphics[width=8.5cm,keepaspectratio]{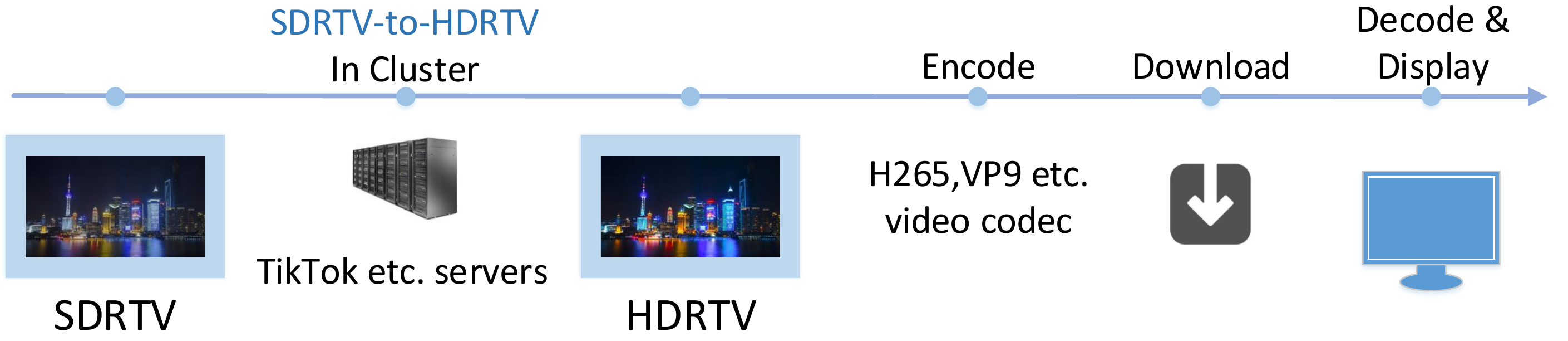}
  \caption{The application scenario of SDRTV-to-HDRTV, after the audio and video media platform converts SDRTV to HDRTV on the cloud server side, it is distributed to the user side through the Internet.}
  \label{streamframework}

\end{figure}


    

There are several formats of HDRTV, and different formats correspond to different EOTF\shortcite{EOTFREF} and metadata. 
The HLG format uses a hybrid log gamma curve as the EOTF, and several other types use a perceptual quantization\shortcite{PQCURVE} (PQ) curve as the EOTF curve.
HDR10 uses static metadata, while the other HDR formats use dynamic metadata.

Compared with SDRTV, HDRTV has significant improvement in visual perception and other aspects, but most of the current video resources are stored in SDR format (historical problems, early video shooting hardware stored video in SDRTV format).
Therefore it is meaningful to build a solution to convert SDRTV to HDRTV. Meanwhile, mainstream streaming platforms have also proposed corresponding solutions to convert SDRTV to HDRTV, and the whole practice process is shown in Fig.\ref{streamframework}.
To facilitate the description, we use SDRTV-to-HDRTV to represent the process of converting SDRTV to HDRTV, which is the same as the previous method \shortcite{refhdrtvnet}.
This paper proposes an end-to-end neural convolution network DSLNet to accomplish the SDRTV-to-HDRTV conversion in this framework.

\begin{figure}[h]
  \centering
  \includegraphics[width=0.45\textwidth]{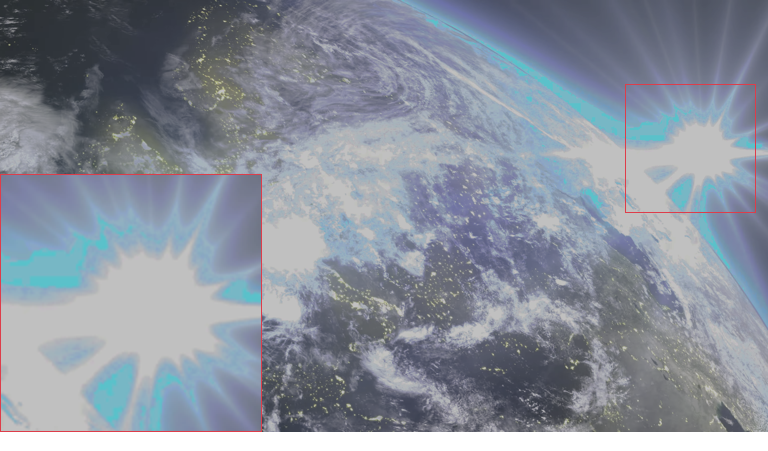}
  \caption{Artifact in converted HDRTV overexposed area.}
  \label{banding}
\end{figure}

Our DSLNet consists of three modules, DMFA, STFM, and LKQE.
First is DMFA. The current frame of the video has a strong correlation with the adjacent frames, and the quality of the current frame can be enhanced by extracting the adjacent frame feature, so we propose a dynamic multi-frame alignment fusion module, DMFA, which can significantly improve the quality of the HDR frames obtained from the conversion.
We are the first method using multi-frame alignment in SDRTV-to-HDRTV.
Next is the STFM module. The previous method \shortcite{refhdrtvnet} processes a single frame of SDR and modulates a single frame by extracting a modulation vector from a single frame.
But this frame-by-frame processing does not take full advantage of temporal information. To improve the conversion quality with temporal information, we propose a spatial-temporal feature modulation (STFM) module, which combines the estimated feature modulation vectors of adjacent frames, and then modulates the current SDR frame more accurately to enhance the converted HDR video quality.
Finally, there is the LKQE module. In the process of converting SDRTV to HDRTV, one of the common problems is that the converted HDRTV has artifact in the highlights area, as shown in the figure \ref{banding}. To this end, a quality enhancement module LKQE with large convolution kernels is designed. The large convolution kernel can effectively extract long-distance image feature information and count large-scale image pixel information, thereby enhancing the image quality of highlighted overexposed areas.

The previous dataset contains only single-frame SDR and single-frame HDR per set and thus cannot be used for the multi-frame method.
For this reason, we propose a dataset called HDRTVMF1K, which can be used for training and evaluating the multi-frame fusion SDRTV-to-HDRTV method.
To evaluate the performance of different methods, we selected three metrics, PSNR, SR-SIM, and $\Delta E_{ITP}$ to evaluate the objective quality, structural similarity, and color fidelity of the proposed method (these metrics support HDR Format).

In summary, our contribution consists of the following four points.
\begin{itemize}
  \item proposes a dynamic large field alignment fusion module that can take advantage of adjacent frame information to improve the quality of converted HDR video frames.
  \item A spatial-temporal feature modulation module is proposed, which can extract spatial-temporal context information to feature modulate the current frame features.
  \item A quality enhancement module with large kernel is proposed, which can effectively alleviate the artifact problem of overexposed areas and improve the objective quality.
  \item The previous dataset is single-frame and cannot be used for training and validation of multi-frame fusion models. We propose a multi-frame SDRTV-to-HDRTV dataset. And the performance of the proposed method is verified on the dataset.  
\end{itemize}
\section{Related Work}

We first introduce the production process of HDRTV and SDRTV.
The RAW data captured by the camera will go through Tone Mapping, Gamut mapping, OETF\shortcite{tonemappref}, Encoder, Decoder, EOTF\shortcite{EOTFREF}, and finally get the optical signal for playback on the monitor, the whole process is shown in Fig.\ref{hdrcapture}.
SDRTV and HDRTV have different settings in the acquisition process, mainly including different tone mapping curves, different EOTF, and OETF, and different gamut mapping, HDRTV has a high dynamic range and a wider color gamut.

\subsection{LDR-to-HDR}

To distinguish SDRTV to HDRTV from LDR to HDR.
We introduce the purpose and method of LDR-to-HDR.
The purpose of the LDR-to-HDR method is to predict the brightness of an image in the linear domain, i.e., the physical brightness of the scene, so HDR images are often saved with 32-bit float precision and commonly used formats are hdr/tif/exr/raw.
Traditional methods estimate the light source density, on top of which the dynamic range is further extended \shortcite{1111,2222,3333,4444}.
Researchers have proposed a deep convolutional neural network-based method \shortcite{liu2020single} to convert LDR images directly to HDR images.
Methods such as HDRCNN\shortcite{HDRCNN,liu2020singleimage,3392403} can restore the overexposed areas of the image.
The method proposed by \shortcite{drhdri,5555,yan2020deep,niu2021hdr} can predict a multi-exposure LDR image pair from a single LDR image, and then synthesize an HDR image based on the predicted multi-exposure image pair.

\subsection{Traditional SDRTV-to HDRTV Conversion}

\begin{figure}[!t]
  \centering
  \subfigure[The production process of HDRTV and SDRTV]{
  \label{hdrcapture}
  \begin{minipage}[h]{\linewidth}
  \centering
  \includegraphics[width=1\textwidth]{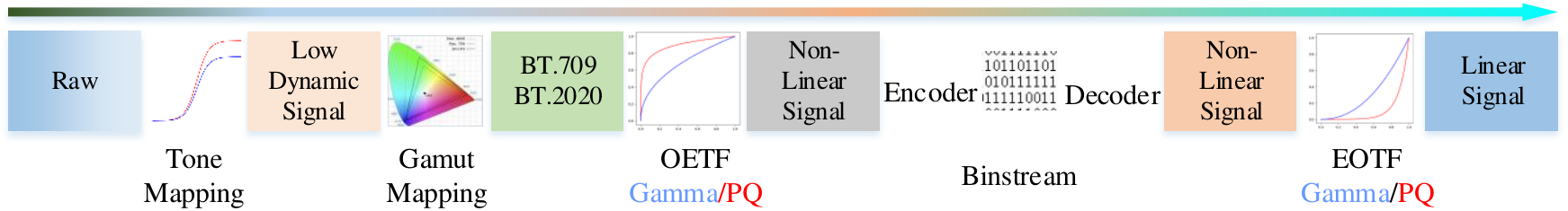}
  \end{minipage}%
  }
  \subfigure[Traditional SDRTV-to-HDRTV solution flow]{
  \label{tradsdr2hdr}
  \begin{minipage}[h]{\linewidth}
  \centering
  \includegraphics[width=1\textwidth]{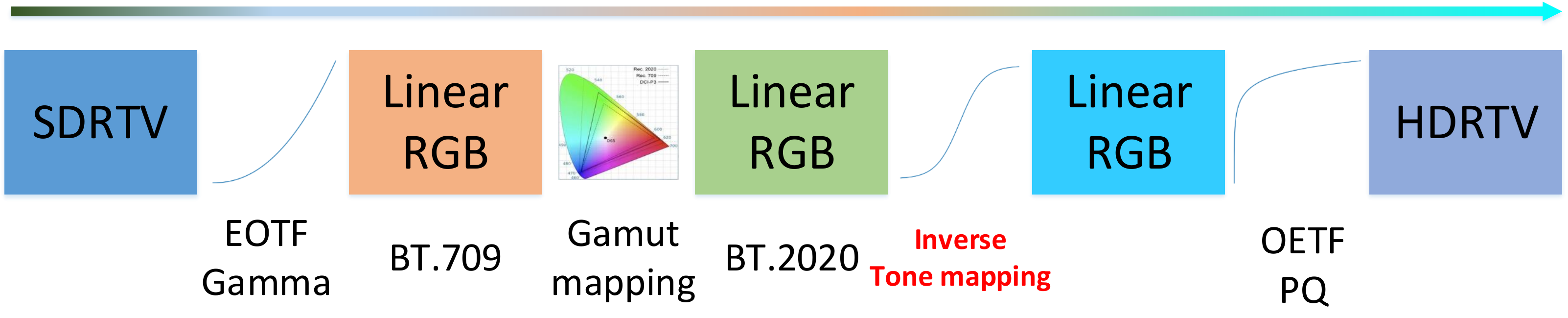}
  \end{minipage}%
  }
  \caption{Analysis of SDRTV-to-HDRTV. 
  (a) The production process of HDRTV and SDRTV.
  (b) Traditional SDRTV-to-HDRTV solution flow.}
  \label{Figure 1 image pipeline}
  \end{figure}

SDRTV-to-HDRTV has high practical value, so many video media manufacturers are doing this and proposing corresponding traditional solutions. The core problem of the traditional solution is to design an effective inverse tone mapping curve to better recover HDR information. The overall framework is shown in Fig.\ref{tradsdr2hdr}.



\subsection{Deep learning SDRTV-to HDRTV Conversion Methods}
The SDRTV-to-HDRTV conversion approach has only emerged in the last two years.
\shortcite{SRITM} proposes a GAN-based architecture that jointly achieves super-resolution and SDTV to HDRTV.
\shortcite{JSIGAN} proposes a hierarchical GAN architecture to accomplish super-resolution and SDRTV to HDRTV.
\shortcite{refhdrtvnet} proposes a method using global feature modulation, local enhancement, and over-exposure compensation, which achieved better performance.

These existing methods perform the SDRTV-to-HDRTV task on a single frame without considering the video frame time correlation.
The method proposed in this paper focuses on the temporal correlation perspective, and the design of the multi-frame fusion SDRTV-to-HDRTV method can better convert SDRTV to HDRTV.

\section{Methodology}

\subsection{Framework}

\begin{figure}[h]
  \centering
  \includegraphics[width=0.48\textwidth]{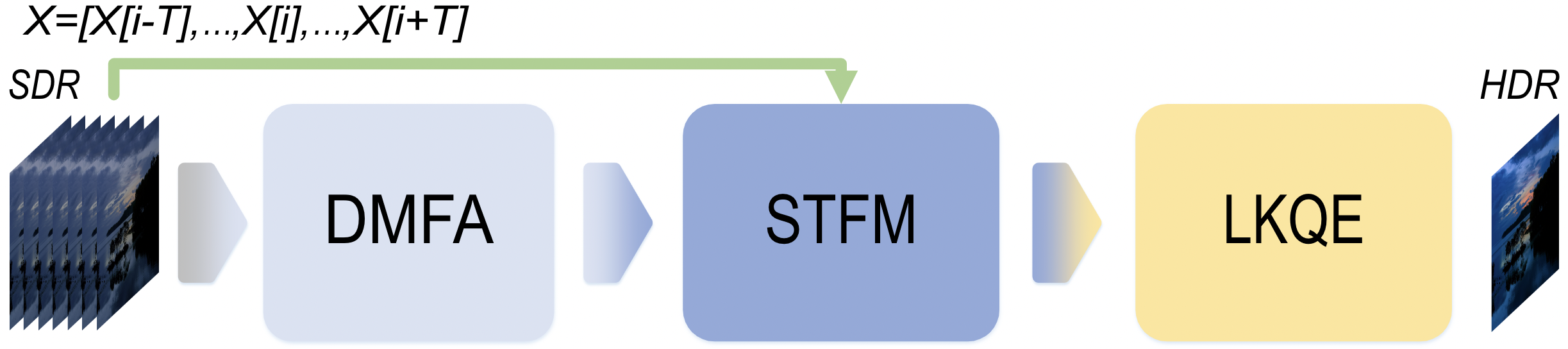}
  \caption{Archotecture of the DSLNet.
  The three modules DMFA, STFM and LKQE accomplish multi-frame alignment fusion, spatial-temporal inverse tone mapping and local quality enhancement, respectively.}
  \label{modelframework}
\end{figure}

Our DSLNet consists of three modules, including \textbf{DMFA} (Dynamic Multi-Frame Alignment), \textbf{STFM} (Spatial-Temporal Feature Modulation) and \textbf{LKQE} (Large Kernel Quality Enhancement). The whole DSLNet framework is shown in Fig.\ref{modelframework}.
DFMA can align multiple frames, extract adjacent frame information, and better reconstruct HDR frames.
STFM can use the spatial-temporal feature modulation vectors extracted from adjacent frames to perform feature modulation on the current frame, and the feature vectors estimated from multiple frames can perform feature modulation more accurately.
LKQE can obtain a very large receptive field and can fix the problem of artifact in overexposed areas in reconstructed HDR video frames.

Given $2T+1$ consecutive SDR frames $[X_{i-T}, X_{i+T}]$, take the center frame $X_i$ as the target frame to be converted, and take the neighboring frame as the reference frame. The input to the model is the target frame $X_i$ and $2T$ adjacent frames, and the output is the enhanced target frame $Y_i$. The objective function is in the formula (\ref{FormulasSDR2HDR}),
where DSLNet is our proposed multi-frame transformation network, and X is the stack of SDR frames.
\begin{equation}
\begin{aligned}
    Y_i & = DSLNet(X) \\
    X & = [X_{i-T},\cdots,X_{i-1},X_{i},X_{i+1},\cdots,X_{i+T}]
\end{aligned}
\label{FormulasSDR2HDR}
\end{equation}
Where $i$ denotes the frame index and T is the maximum number of reference frames.
In the following chapters, we will provide a detailed analysis of the motivation and architecture of each module.

\subsection{DMFA - Dynamic Multi Frame Alignment}

\begin{figure}[h]
  \centering
  \includegraphics[width=0.48\textwidth]{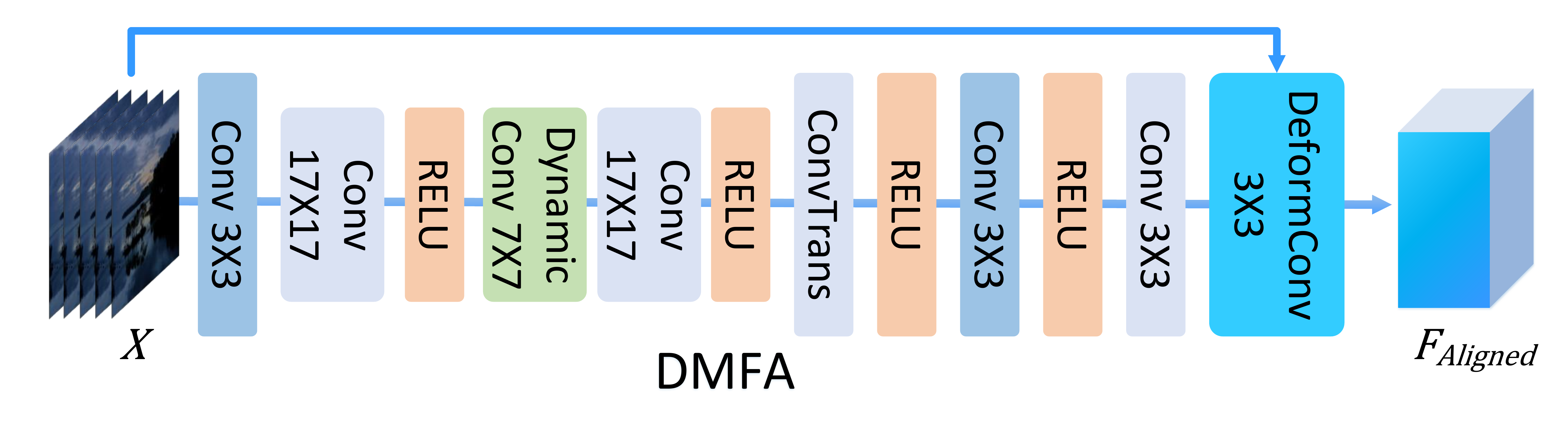}
  \caption{Architecture of Dynamic Multi-Frame Alignment Module(DMFA).}
  \label{dmfamodule}
\end{figure}

Using adjacent frame information is valuable for image recovery/enhancement. Early multi-frame fusion methods \shortcite{AndrejKarpathy2014LargeScaleVC} directly concat multiple frames and then use convolution to extract features, which are simple but do not accomplish spatial-temporal alignment.
In order to better accomplish the alignment of frame details, we propose a dynamic multi-frame alignment module DMFA.
The architecture of DMFA is shown in Fig.\ref{dmfamodule}.

\subsubsection{Motivation of DMFA}
There is a very large temporal correlation between adjacent frames of SDR video, so it is helpful to use the adjacent frame information for the current SDR frame to the HDR frame.
Adjacent frames are crucial for HDR frame conversion, but due to viewpoint, motion, and video compression issues, adjacent video frames need to be aligned before they can be accurately utilized.
We propose the corresponding DMFA module, which can capture temporal displacement such as large-scale motion information to complete video frame alignment fusion.

\subsubsection{Architecture of DMFA}

We present the detailed architecture of DMFA here.
To enable efficient multi-frame alignment, we design a multi-frame alignment module with a dynamic large convolutional kernel for offset estimation.
For the input SDR video frame sequence $X$, a feature alignment fusion of $X$ is required.
We use the deformable convolution \shortcite{deformableconvpaper} method to perform feature alignment.
In order to estimate the offset more accurately, our proposed DMFA module is designed with a dynamic offset estimation module (LDOE) for the large convolution kernel.
LDOE extracts non-local motion information through the large convolution kernel and can accomplish long-range motion alignment. In addition, the dynamic convolution adaptively adjusts the estimated offset according to the video frame sequence, which can eventually obtain a more accurate offset.

We introduce the detailed composition of the LDOE module. For the input SDR video frame sequence $X$, it first goes through a $3 \times 3 $ ordinary convolution (stride set to 2) and RELU activation function to extract the multi-frame nonlinear features.
Then it goes through a $17 \times 17 $ Depthwise convolution, which helps to capture long-range features and thus enables to estimate large scale motion information.
The next step is a $7 \times 7$ dynamic convolution \shortcite{idynamic}, which allows adaptive processing of features based on the image content (the convolution kernel weights in dynamic convolution are predicted from the current image, which enables content-adaptive processing of the image and captures more complex and variable functional mapping relationships compared to static convolution).
Next, the offset feature map $X_{offset}$ with the same resolution as the original image is obtained after $17 \times 17 $ Depthwise convolution, RELU, transposed convolution, and RELU.
Finally, $X_{offset}$ is input to the deformable convolution with $X$ to obtain the aligned spatial-temporal features $X_{Aligned}$.

\subsection{STFM - Spatial Temporal Feature Modulation}

We propose a novel spatial-temporal feature modulation module (STFM), which consists of three parts: feature modulation vector estimation module, feature modulation, and residual block.
Our innovative design of the spatial-temporal feature modulation estimation module, which combines adjacent frames to estimate the feature modulation vector of the current frame, can be used for more accurate feature modulation and ultimately higher quality HDR video frames.
Our proposed spatial-temporal feature modulation module consists of three main parts.

1. Branch 1. A branch of a conditional network that predicts spatial-temporal feature modulation vectors.

2. Branch 2. The spatial-temporal feature modulation branch modulates the input feature with the predicted spatial-temporal feature modulation vector.

3. Branch 3. A parallel depthwise separable residual block with large cores is constructed to extract features.

\subsubsection{Motivation of STFM}

The previous SDRTV-to-HDRTV method \shortcite{refhdrtvnet} converts single-frame SDR frames to HDR frames, using a modeling approach that uses feature mapping of a single frame.
Estimating the feature mapping function by a single frame is relatively simple. It does not consider the video's temporal correlation, so the feature mapping capability is limited. In addition, for video, using single frames for processing can lead to problems such as flicker.
The adjacent frames in the video correlate with the current frame, so the adjacent frame information can help estimate the feature mapping function better and thus get better conversion of the obtained HDR video, which is our motivation to design the temporal feature modulation.

\subsubsection{Architecture of STFM}

\begin{figure}[h]
  \centering
  \includegraphics[width=0.49\textwidth]{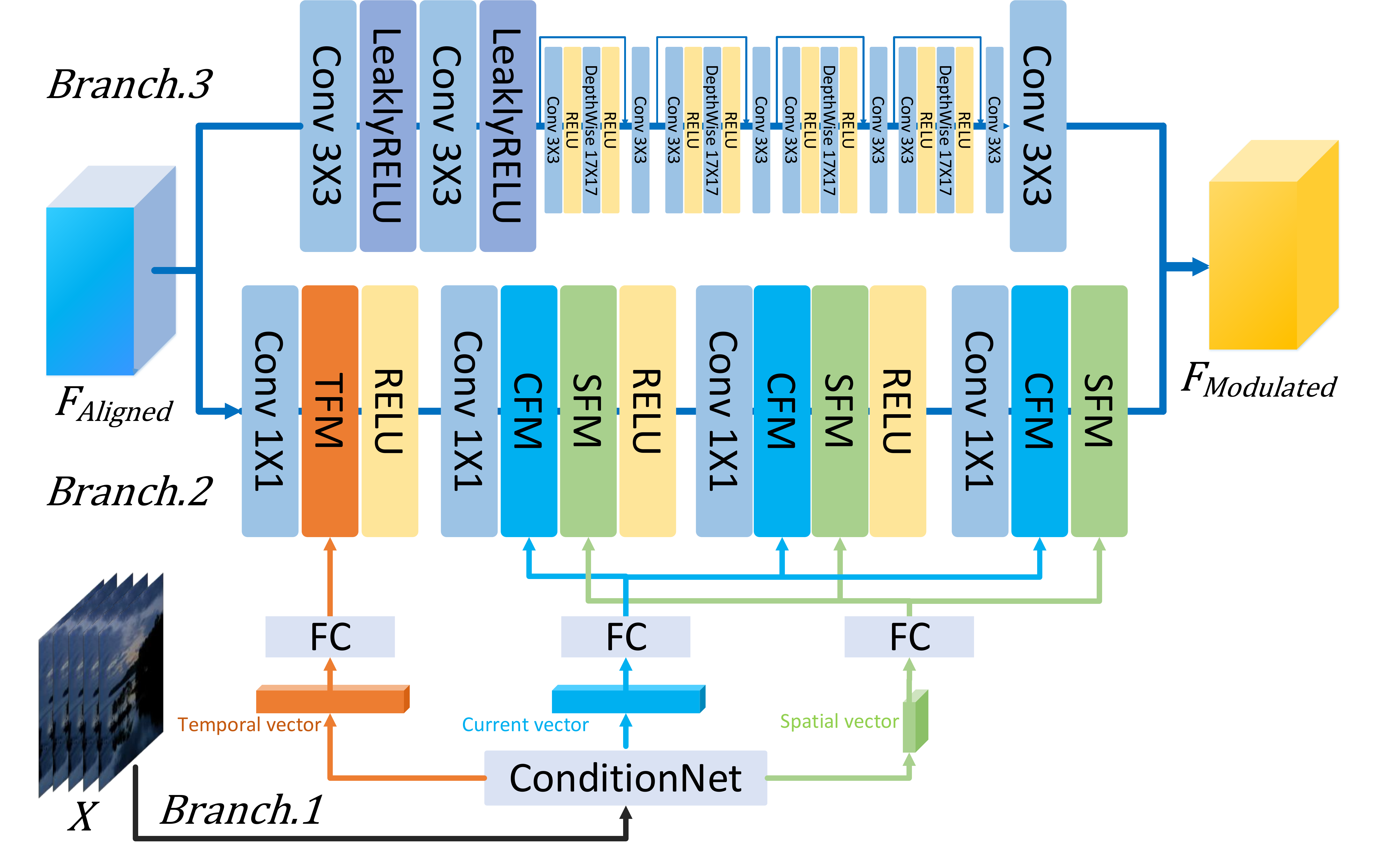}
  \caption{Architecture of Spatial Temporal Feature Modulation(STFM).}
  \label{stfm}
\end{figure}

The inputs to the STFM module are the aligned features $F_{Aligned}$ and the input frame sequence $X$.
The STFM contains three branches, the conditional network, the modulation branch, and the residual block branch.
The architecture of STFM is shown in Fig.\ref{stfm}.

\begin{figure}[h]
  \centering
  \includegraphics[width=0.45\textwidth]{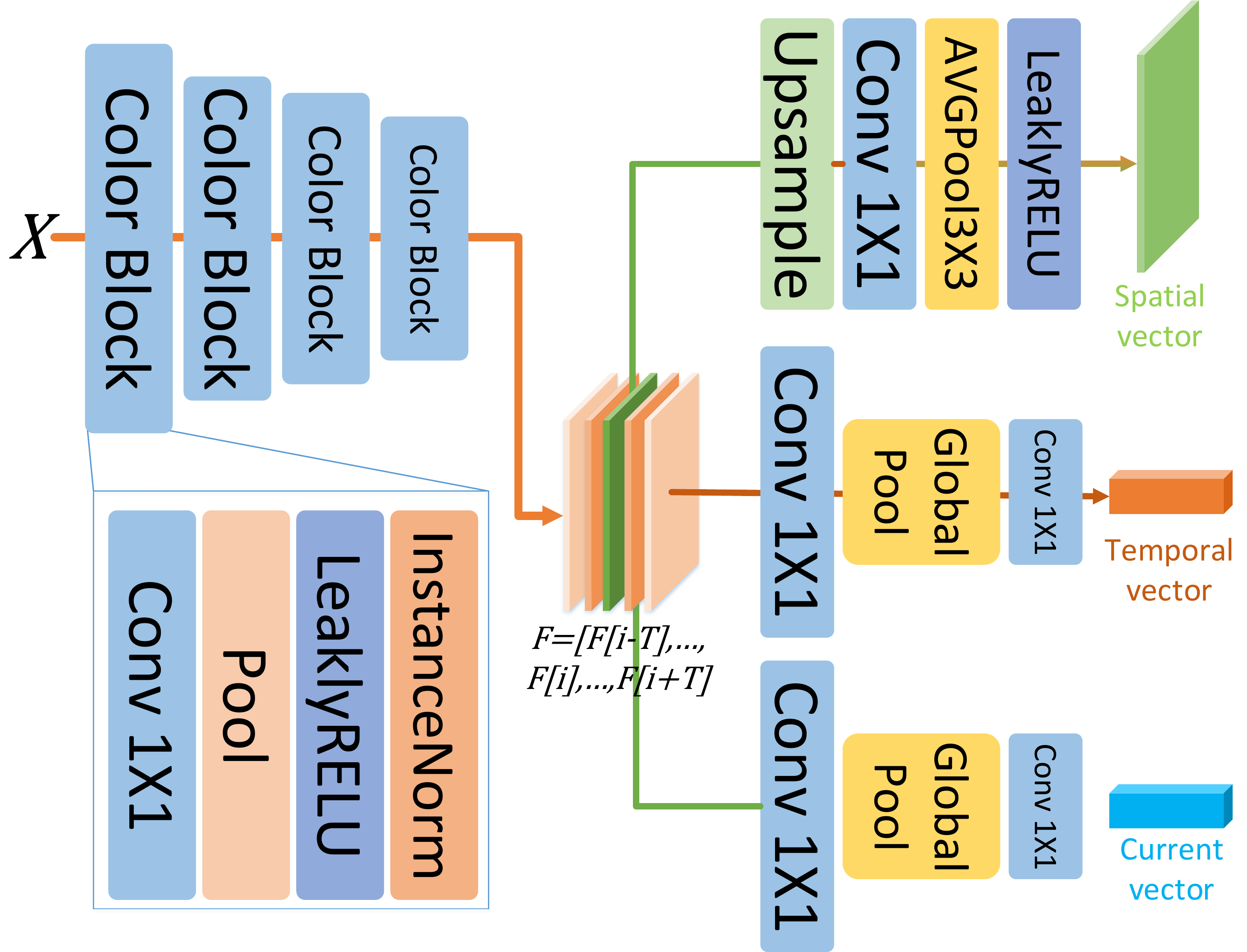}
  \caption{Architecture of ConditionNet.}
  \label{conditionnet}
\end{figure}

\textbf{STFM: Branch 1. ConditionNet}.
The architecture of the condition network is shown in Fig.\ref{conditionnet}.
In the condition network, the input video frame sequence $X$ is fed into the serial 4 ColorBlocks, each of which has the structure given in Eq. \ref{CB}, the feature $F$ of each frame in the sequence is obtained, $F=[F_{i-T},\cdots,F_{i-1},F_{i},F_{i+1},\cdots,F_{i+T}]$.
The current frame feature $F_{i}$ is input to the spatial modulation vector estimation module to obtain the final spatial feature modulation vector $V_{SM}$. The specific calculation procedure of SME is shown in Eq. \ref{SME}.

\begin{equation}
  \begin{aligned}
    &ColorBlock(X) \\
    &=  Conv1\times1(Pool(ReLU_{leaky}(Norm(X))))
  \end{aligned}
  \label{CB}
\end{equation}

\begin{equation}
\begin{aligned}
  & V_{SM} = SME(F_i) \\
     & = Up(Conv1\times1(Pool_{AVG}(ReLU_{leaky}(F_i))))
\end{aligned}
\label{SME}
\end{equation}
The multi-frame features $F$ extracted from the frame sequence are all input to the temporal modulation vector estimation module TME to obtain the temporal feature modulation vector $V_{TM}$. The calculation procedure of TME is shown in Eq. \ref{TME}.
\begin{equation}
\begin{aligned}
    & V_{TM} = TME(F) \\
    & = Conv1\times1(Pool_{Global}(Conv1\times1(F)))
\end{aligned}
\label{TME}
\end{equation}
The feature $F_{i}$ are input to the current frame modulation vector estimation module CME to obtain the final current frame feature modulation vector $V_{CM}$. The specific calculation procedure of CME is given in Eq. \ref{CME}.
\begin{equation}
\begin{aligned}
    & V_{CM} = CME(F_i) \\
   & = Conv1\times1(Pool_{Global}(Conv1\times1(F_i)))
\end{aligned}
\label{CME}
\end{equation}

\textbf{STFM: Branch 2. Modulation}.
In the modulation branch, it first goes through a 1 $\times$ 1 convolution, followed by TFM.

\textbf{TFM-Temporal Feature Modulation}.
The TFM is the temporal feature modulation module that models the adjustment process of adjacent frames to the current frame.
The input to the TFM is the feature $F_P$ output from the previous layer with the temporal modulation vector $V_{TM}$, $V_{TM}$ consisting of $V_{TMA}$ and $V_{TMB}$, the parameters for translation and scaling during the modulation, respectively.
The formal description of the TFM is given in Eq. (\ref{tfmform}), the process of feature modulation by $V_{TMA}$ scaling the input features $F_P$ and $V_{TMB}$ panning the input features by the modulation vector predicted from adjacent frames to the current frame is the temporal feature modulation.
\begin{equation}
\begin{aligned}
     Y & = TFM(F_P,V_{TMA},V_{TMB}) \\\
   & = F_P*V_{TMA}+V_{TMB}
\end{aligned}
\label{tfmform}
\end{equation}


After completing the temporal feature modulation, the adjustment of adjacent frames to the current frame is completed.
Next, the frame adaptive feature modulation, including CFM and SFM, is required.
The final modulation branch is formed by repeating three times the series RELU, Conv1$\times$1, CFM, and SFM to obtain the modulated feature $F_{Mod}$, the architecture of which is shown in Fig.\ref{stfm}.

\textbf{Current Feature Modulation}.CFM is the current frame feature modulation, which uses the vector $V_{CM}$ estimated from the current frame to modulate the current frame. This part models the function of frame adaptive adjustment.

\textbf{Spatial Feature Modulation}.SFM is the current frame spatial feature modulation. This part uses the vector $V_{SM}$ estimated from the current frame to modulate the features of the current frame.
Unlike CFM, $V_{SM}$ is not a single vector, the number of vectors of $V_{SM}$ is consistent with the resolution, and there is a corresponding modulation vector for each pixel position so that the local adaptive adjustment of the frame can be accomplished.
$V_{CM}$ and $V_{SM}$ are composed of two parts, including $V_{CMA}$ and $V_{CMB}$,$V_{SMA}$ and $V_{SMB}$, which are the scaling and panning parameters of the modulation process, respectively.
The CFM and SFM are calculated as shown in Eq. (\ref{scfmform})

\begin{equation}
\begin{aligned}
 Y & = CFM(F_P,V_{CMA},V_{CMB}) \\\
& = F_P*V_{CMA}+V_{CMB}\\
 Y & = SFM(F_P,V_{SMA},V_{SMB}) \\\
& = F_P*V_{SMA}+V_{SMB}
\end{aligned}
\label{scfmform}
\end{equation}

\begin{figure*}[h]
  \centering
  \includegraphics[width=1.0\textwidth]{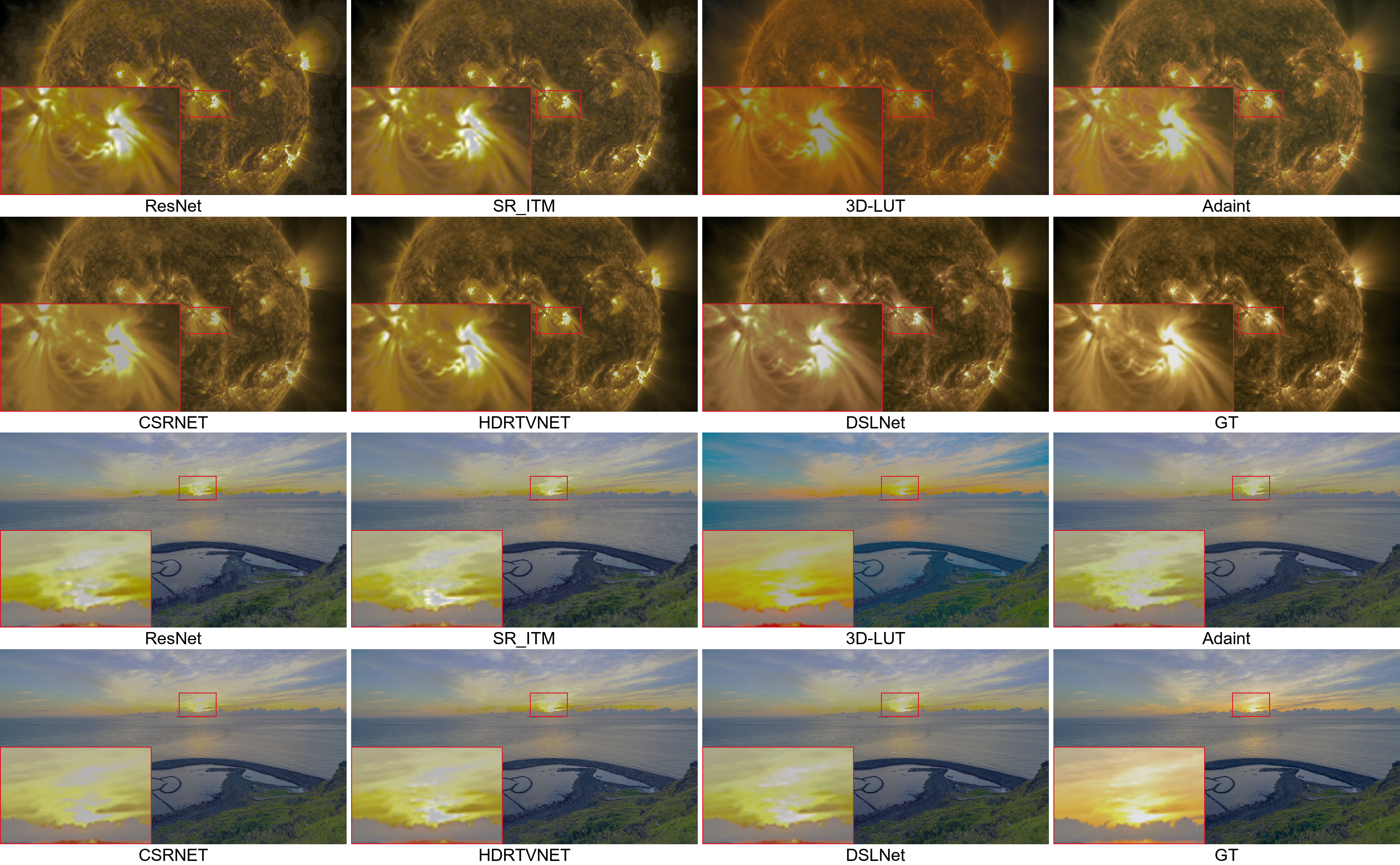}
  \caption{Comparison of visual quality by different SDRTV-to-HDRTV methods 
  (playing HDR video frames on an SDR monitor will appear grayed out, we do not do additional tone mapping to preserve all the information of HDR frames).}
  \label{visimage}
\end{figure*}

\begin{table*}[h]
  \begin{spacing}{1.0}
  \centering
  \begin{tabular}{m{6.2cm}<{\centering}m{1.5cm}<{\centering}m{1.5cm}<{\centering}m{1.5cm}<{\centering}m{1.5cm}<{\centering}m{1.5cm}<{\centering}m{2.5cm}<{\centering}}
  \toprule
  Methods & Params$ \downarrow$ & PSNR$ \uparrow$  & SR-SIM$ \uparrow$  & $\Delta E_{ITP} \downarrow$    \\ 
  \midrule
  ResNet\shortcite{he2016identity}ECCV16 & 1.37M & 33.62   & 0.9913 & 12.72   \\
  CSRNET\shortcite{he2020conditional}ECCV20 & 36K & 34.53  & 0.9946 & 13.23   \\
  Ada-3DLUT\shortcite{zeng2020learning}TPAMI  & 0.59M & 26.15    & 0.9846 & 37.99  \\
  AdaInt\shortcite{yang2022adaint}CVPR2022    & 0.62M & 30.99    & 0.9931 & 22.33  \\
  SR-ITM\shortcite{SRITM}ICCV21               & 2.87M & 33.99    & 0.9893 & 12.85   \\
  HDRTVNET\shortcite{refhdrtvnet}ICCV21       & 1.41M & 34.66    & 0.9938 & 12.38   \\
  \textbf{DSLNet}(Proposed)                   & 0.93M & 35.28    & 0.9951 & 11.13  \\ 
  \bottomrule

  \end{tabular}
  \end{spacing}
  \caption{Quantitative results of different methods on the parameters, PSNR, SR-SIM and $\Delta E_{ITP}$. $\uparrow$ or $\downarrow$  means that larger or smaller is better.}
  \label{Quantitativetable}
\end{table*}

\textbf{STFM: Branch 3. Residual block}.

Our motivation for designing the residual branch is to transfer the unmodulated features directly to the end after convolution so that our modulated branch only needs to learn the residual part.
Such a design allows unmodulated features to be transferred directly to the end, and the model can adaptively fuse modulated and unmodulated features.
Eventually, more features can be preserved, resulting in a better conversion of SDR video to HDR video.

We present here the specific architecture of the residual branch.
For the input $F_{Aligned}$ features, they first go through Conv $3\times3$ and leakyReLU twice.
Then the residual features $F_{skip}$ are obtained by passing through four large kernel residual blocks LKRB and Conv $3\times3$ in series. Each residual block LKRB contains Conv 3$\times$3, ReLU, DepthWise 17$\times$17 and ReLU in series. The computation procedure is shown in Fig.\ref{DW1}.
\begin{small}
\begin{equation}
\begin{aligned}
F_{skip} & = Conv 3\times3(LKRB(F_{Aligned})) \\\
LKRB(X) & = Conv 3\times3 (ReLU(DW17\times17 (ReLU(X))))
\end{aligned}
\label{DW1}
\end{equation}
\end{small}
The output $F_{Modulated}$ of STFM is obtained by adding $F_{Mod}$ with $F_{skip}$.
Our proposed spatial-temporal feature modulation module STFM can modulate the current frame features more accurately, thus improving the quality of the converted HDR video.

\subsection{LKQE - Large Kernel Quality Enhancement}

To further enhance the quality of HDR frames, we designed a quality enhancement module LKQE with a large core.
The large convolutional kernel captures long-range information to enhance the quality of the HDR video obtained by conversion.

\subsubsection{Architecture of LKQE}

The architecture of our Large Kernel Quality Enhancement (LKQE) module is simple.
The modulated feature $F_{Aligned}$ is input to the LKQE module, first through Conv 3$\times$3 and RELU.
Next, four large kernel residual blocks $LKRB$ are passed in turn. The architecture of each residual block is the same as the upper half of the parallel residual branch in Figure \ref{stfm}.

\section{Experiment}

\subsection{Dataset}

Since the training and testing of multi-frame models requires multi-frame SDRTV-to-HDRTV datasets, the dataset used by \shortcite{refhdrtvnet} is only used for single-frame SDRTV-to-HDRTV.
Therefore we produced a multi-frame SDRTV-to-HDRTV dataset.
We collected 46 HDR videos of HDR10 standard with a resolution of 2160$\times$3840 from YouTube, and each HDR video has a corresponding SDR video.
All videos are encoded with PQ curve and BT.2020 color gamut.
42 pairs of videos are used for training, and four pairs of videos are used for testing.
We first perform scene segmentation for each video and then take ten consecutive frames for each scene. We perform cropping operations in the training set (the test set uses full-resolution frames for testing).
The final training set contains 18790 sets of data, and each set contains 10 corresponding frames of HDR and 10 frames of SDR.
The test set contains 74 sets of video sequences, and each set contains 10 frames of HDR and 10 frames of SDR.

\subsection{Training Details}

During the training of the model, we optimize DSLNet using L1 as the loss function.
Using the Adam optimizer, the initial learning rate is set to 0.0005, and the learning rate is set to 1/2 of the current learning rate every 50,000 iterations.
After the 150,000th iteration, the learning rate is set to 1/2 of the current learning rate for every 40,000 iterations; the total number of iterations is set to 350,000.

\subsection{Evaluation Metrics}

To effectively evaluate the quality of the generated HDR videos, we used three evaluation metrics, PSNR, SR-SIM, and $\Delta E_{ITP}$, for a comprehensive evaluation.
SR-SIM is an image similarity metric, and SR-SIM has good evaluation performance for HDR videos.
We introduced $\Delta E_{ITP}$ to measure the color fidelity, and $\Delta E_{ITP}$ is designed for HDRTV.

\subsection{Quantitative Results}

Table.\ref{Quantitativetable} lists the quantitative comparison results.
It can be seen that our DSLNet achieves the best conversion performance.
Specifically, the PSNR of our DSLNet is at least 0.62 dB higher than that of other methods.
On SR-SIM, our DSLNe outperforms the other methods by at least 0.0005.
On $\Delta E_{ITP}$, DSLNet is at least 1.25 lower than other methods, which indicates that our DSLNet gains a great improvement in color fidelity. All these results show that our DSLNet has a significant improvement in objective quality compared to the previous state-of-the-art methods.

\subsection{Qualitative Results}

Here we visualize the results of the different methods by playing the 16Bit image directly, as this preserves all the video frame information to the maximum extent.
As we can see from Fig.\ref{visimage}, our DSLNet produces a better visualization than the competitor's.
For example, in the overexposed region in row 1, we observe that our DSLNet achieves a better subjective quality.

\subsection{Ablation study}

\subsubsection{Ablation study on each module}

\begin{table}[h]
  \centering
  \begin{tabular}{m{0.1cm}<{\centering}m{0.1cm}<{\centering}m{0.1cm}<{\centering}m{0.1cm}<{\centering}m{0.1cm}<{\centering}m{0.1cm}<{\centering}m{0.1cm}<{\centering}m{0.1cm}<{\centering}m{1.cm}<{\centering}m{0.8cm}<{\centering}m{0.8cm}<{\centering}} 
  \toprule
   M0    &   M1    &  M2     &  M3      &  M4    &  M5     &  M6     &  M7   &  PSNR$\uparrow$  &  SR-SIM$\uparrow$ &  $\Delta E_{ITP}$$\downarrow$  \\ 
  \midrule
  \Checkmark        & \XSolidBrush       & \XSolidBrush       & \XSolidBrush        & \XSolidBrush      & \XSolidBrush       & \XSolidBrush       & \XSolidBrush     & 33.80   & 0.9913 &  13.07 \\
  \Checkmark        & \Checkmark         & \XSolidBrush       & \XSolidBrush        & \XSolidBrush      & \XSolidBrush       & \XSolidBrush       & \XSolidBrush     & 34.86   & 0.9943 &  12.65 \\
  \Checkmark        & \Checkmark         & \Checkmark         & \XSolidBrush        & \XSolidBrush      & \XSolidBrush       & \XSolidBrush       & \XSolidBrush     & 34.99   & 0.9940 &  12.58 \\
  \Checkmark        & \Checkmark         & \Checkmark         & \Checkmark          & \XSolidBrush      & \XSolidBrush       & \XSolidBrush       & \XSolidBrush     & 35.01   & 0.9943 &  12.39 \\
  \Checkmark        & \Checkmark         & \Checkmark         & \Checkmark          & \Checkmark        & \XSolidBrush       & \XSolidBrush       & \XSolidBrush     & 35.04   & 0.9944 &  12.67 \\
  \Checkmark        & \Checkmark         & \Checkmark         & \Checkmark          & \Checkmark        & \Checkmark         & \XSolidBrush       & \XSolidBrush     & 35.07   & 0.9945 &  12.44 \\
  \Checkmark        & \Checkmark         & \Checkmark         & \Checkmark          & \Checkmark        & \Checkmark         & \Checkmark         & \XSolidBrush     & 35.13   & 0.9948 &  11.71 \\
  \Checkmark        & \Checkmark         & \Checkmark         & \Checkmark          & \Checkmark        & \Checkmark         & \Checkmark         & \Checkmark       & 35.28   & 0.9951 &  11.13 \\

  \bottomrule
  \end{tabular}
  \caption{Ablation study to verify the validity of the four modules, M0,M1,M2,M3,M4,M4,M5,M6,M7 refer to Baseline,
  Single Frame Feature Modulation,
  Parallel Large Kernel Depthwise Block,
  Temporal Feature Modulation,
  Spatial Feature Modulation,
  Large Kernel Depthwise Alignment,
  Large Kernel Depthwise Quality enhancement,
  Dynamic Alignment. 
  The addition of each part can bring about the improvement of PSNR and other indicators, which proves that each module is indeed effective. }
  
          \label{abl0}
\end{table}

In order to verify the effectiveness of each component of the proposed method, we performed a full-scale module ablation experiment.
Here we add each module in turn so that we can derive the performance gain from each module.
We choose a single frame Resnet as our baseline and add the modules sequentially on top of it.
For ease of representation, we use symbolic shorthand. Use M0 for baseline.
Add each module in turn on top of M0 (feature modulation M1, parallel large-core depth-separable convolutional residual block M2,
Temporal feature modulation M3, spatial feature modulation M4, large kernel depth-separable convolutional motion estimation block M5,
large kernel mass enhancement block M6, dynamic convolutional motion estimation M7)
Table.\ref{abl0} shows the gain from the addition of each module.

\subsubsection{Ablation study on Multi-frame Fusion}

To verify the design validity of our multi-frame fusion model, we designed another multi-frame Resnet model by inputting adjacent frames as well as the current frame together into the multi-frame Resnet model.
From Table.\ref{ablmresnet}, we can see that the HDR video quality obtained by conversion is higher with a smaller number of parameters in our model due to the sophisticated design of the proposed method.
The design effectiveness of the proposed method is proved.

\begin{table}[h]
  \centering
  \begin{tabular}{m{2.0cm}<{\centering}m{0.8cm}<{\centering}m{0.8cm}<{\centering}m{1.3cm}<{\centering}m{1.1cm}<{\centering}m{1.1cm}<{\centering}}
    \toprule
    Model   & Param & PSNR  & SR-SIM & $\Delta E_{ITP}$ \\
    \midrule
    DSLNet  & 0.93M & 35.28  & 0.9951 & 11.13 \\
    MResnet & 1.38M & 33.89  & 0.9906 & 12.31 \\
    \bottomrule
  \end{tabular}
  \caption{The proposed method is compared with the multi-frame Resnet(MResnet).}
  
  \label{ablmresnet}
\end{table}

\subsubsection{Ablation study on Alignment}

We ablate the alignment module in the proposed model to verify the necessity of alignment in this task.
A stacked block of normal convolutional modules is used instead of the deformable convolutional alignment module on top of DSLNet.
It can be seen from Table.\ref{abldeformable} that the performance of the model degrades significantly after dropping the alignment module.

\begin{table}[h]
  \centering
  \begin{tabular}{m{2.0cm}<{\centering}m{0.8cm}<{\centering}m{0.8cm}<{\centering}m{1.3cm}<{\centering}m{1.1cm}<{\centering}m{1.1cm}<{\centering}}
    \toprule
    Model     & Params & PSNR  & SR-SIM & $\Delta E_{ITP}$ \\
    \midrule
    DSLNet    & 0.93M  & 35.28  & 0.9951 & 11.13 \\
    W/o Align & 1.03M  & 34.28  & 0.9935 & 14.24 \\
    \bottomrule
  \end{tabular}
  \caption{Ablation experiments on the alignment module.}
  \label{abldeformable}
\end{table}

\subsubsection{Ablation study on Feature Modulation}

To verify the effectiveness of feature modulation in the model, we use stacked convolution instead of the original modulation module, and as can be seen from Table.\ref{ablmodulation}, the performance of the model without the modulation module drops significantly, verifying the necessity of the modulation module.

\begin{table}[h]
  \centering
  \begin{tabular}{m{2.5cm}<{\centering}m{0.8cm}<{\centering}m{0.8cm}<{\centering}m{1.2cm}<{\centering}m{0.9cm}<{\centering}m{0.9cm}<{\centering}}
    \toprule
    Model           & Params & PSNR  & SR-SIM & $\Delta E_{ITP}$ \\
    \midrule
    DSLNet          & 0.93M & 35.28  & 0.9951 & 11.13 \\
    W/o Modulation  & 1.03M & 33.36  & 0.9925 & 14.04 \\
    \bottomrule
  \end{tabular}
  \caption{Ablation experiments on the Modulation module.}
  \label{ablmodulation}
\end{table}

\section{Conclusion}

In this paper, we develop a multiframe fusion space-time network (DSLNet) for SDRTV-to-HDRTV conversion.
Our approach is the first to introduce multi-frame information fusion in the SDRTV-to-HDRTV process.
For better reconstruction of details, we design a multi-frame aligned feature fusion module.
For better color feature mapping, the proposed method designs a spatial-temporal feature modulation module, which can combine spatial-temporal feature information for feature modulation.
We further extend the large kernel depth separable convolution in our model, which can enhance the quality of the HDR video obtained by conversion.
We also construct a large-scale dataset for training and testing the performance of the multi-frame SDRTV-to-HDRTV method.
Experimental results on the test set show that our DSLNet outperforms previous state-of-the-art methods on the SDRTV-to-HDRTV task.

\bibliography{aaai23}

\end{document}